\documentclass[12pt]{article}
\usepackage{amssymb}
\usepackage{graphicx}
\usepackage[cp1251]{inputenc}
\usepackage{rotating}

\begin{document}
 \noindent {\footnotesize\it Astronomy Reports, 2024, Vol. 68, No. 3, pp. 247--256}
 \newcommand{\dif}{\textrm{d}}

 \noindent
 \begin{tabular}{llllllllllllllllllllllllllllllllllllllllllllll}
 & & & & & & & & & & & & & & & & & & & & & & & & & & & & & & & & & & & & & &\\\hline\hline
 \end{tabular}

  \vskip 0.5cm
  \bigskip
 \bigskip
\centerline{\large\bf Is the young star association $\epsilon$ Cha double?}
 \bigskip
 \bigskip
  \centerline { %  DOI: 10.1134/S1063772924700264
   V. V. Bobylev\footnote [1]{bob-v-vzz@rambler.ru},  A. T. Bajkova}
 \bigskip
 \centerline{\small\it Pulkovo Astronomical Observatory, Russian Academy of Sciences, St. Petersburg, 196140 Russia}
 \bigskip
 \bigskip
{{\bf Abstract}--The kinematics of the $\epsilon$~Cha young stellar association close to the Sun has been studied based on a list of candidate stars from the Dickson--Vandervelde work. The working sample consists of 26 stars with parallaxes, proper motions from the Gaia~DR3 catalog and radial velocities taken from literary sources. The orbits of the stars back to the past were constructed, and the moment when the association had a minimum spatial size was determined, as well as an analysis of the dependencies of the velocities $U,V,W$ on the coordinates $x,y,z$ was carried  out. It is shown that the initial sample is divided into two parts with different kinematic properties. The first sample included 9 stars. Based on the construction of the orbits of these 9 stars, an age estimate of $t=4.9\pm0.8$~million years was obtained. An expansion coefficient in the $xz$ plane with the value $K_{xz}=135\pm19$~km/s/kpc was also found for them, on the basis of which another age estimate $t=7.2\pm1.0$~million years was obtained. The second sample included 17 stars. The construction of their orbits gave an estimate of age $t=0.2\pm0.3$~million years, and based on the gradient $\partial W/\partial z=707\pm248$~km/s/kpc, a second estimate of their age $t=1.4\pm0.5$~million years was obtained. This suggests that the $\epsilon$~Cha association either consists of two groupings of different ages, or a younger one arose as a result of a recent outbreak of star formation within a common star system. The question of the gravitational connection of the groupings has not been considered in the framework of this work.
 }

\bigskip
\section{INTRODUCTION}
A number of small associations and compact moving groups are known near the Sun, consisting of very young (less than about 10 million years old) stars. For example, $\beta$~Pic, TW~Hya, $\eta$~Cha, $\epsilon$~Cha, which are located in the fourth galactic quadrant and are closely related [2] with the extensive Sco-Cen OB association.

The subject of this work is the compact stellar association $\epsilon$~Cha, which is sometimes referred to as the young open star cluster Feigelson~1. Actually the star $\epsilon$~Cha is a visual binary (WDS J11596--7813AB) with a main component of spectral type B9V. The remaining members of the association are, as a rule, low-mass T Tauri stars. Association studies $\epsilon$~Cha is the subject of a large number of publications, where the authors are interested in the stellar composition, kinematics, as well as photometric and spectral characteristics of members and candidates for members of the association [1--12].

Already in work [2] based on integration back into the past, the common origin of such structures as LCC (Lower Centaurus Crux), UCL (Upper Centaurus Lupus), US (Upper Scorpius), TW~Hya, $\eta$~Cha and $\epsilon$~Cha at time approx $-$10 million years (Myr). Moreover, to calculate the average association trajectory $\epsilon$~Cha, only 4 stars were used. Also using the method of integrating orbits over a small number of association member stars $\eta$~Cha and $\epsilon$~Cha at work, [5] their age was estimated at 6.7 million years.

Thanks to the works [1, 3--9] number of probable members of the association $\epsilon$~Cha increased significantly.
This made it possible, for example, to estimate the age of an association by fitting to a suitable theoretical
isochrone using a large number of stars. Thus, using photometric data from the Gaia~DR2 catalog
[13] for more than 20 stars in recent work, [1] this method was used to estimate $5^{+3}_{-2}$~million years.

Stars of the association $\epsilon$~Cha are located at an average distance of 102 pc from the Sun. Due to its
proximity to the Sun, many kinematic and photometric parameters measured with high precision are available
to study this association. Trigonometric parallaxes from the Gaia~DR3 catalog are known for almost
all candidate stars [14], obtained with relative errors less than 5\%.

To calculate the spatial velocities of stars, measurements of their radial velocities are important. Despite the presence of radial velocities in the Gaia~DR3 catalogue, the use of more accurate radial velocities of stars obtained from ground-based observations remains relevant at present. This is especially true if there is a large percentage of doubles in the association $\epsilon$~Cha (40\% according to [1]). For this purpose, individual
spectroscopic measurements of individual stars-members of the association were used $\epsilon$~Cha. In some cases, [11] extensive catalogs of radial velocities were used, for example, RAVE (The Radial Velocity Experiment [15]). Large database of association stars Cha, which includes measurements of their trigonometric parallaxes, proper motions and radial velocities, was created in the work [1].

Youth association $\epsilon$~Cha and its proximity to the Sun allows us to perform unique work, for example, to
carry out direct telescopic observations to search for stars with protoplanetary disks [12], search for low-mass
brown dwarfs of low luminosity, etc.

The purpose of this work is to study the kinematics of association $\epsilon$~Cha and kinematic estimate of its age.
The method consists in constructing the orbits of stars in the past at a given time interval, and an assessment
of the moment when the stellar grouping had a minimum
spatial size. It is also interesting to estimate kinematic age based on the association broadening effect.

\section{METHODS}
We use a rectangular coordinate system centered on the Sun, where the axis $x$ directed towards the
galactic center, axis $y$~--- towards the galactic rotation and axis $z$~--- to the north pole of the Galaxy. Then
$x=r\cos l\cos b,$ $y=r\sin l\cos b$ and $z=r\sin b,$ where $r=1/\pi$ is the heliocentric distance of the star in
kpc, which we calculate through the trigonometric parallax of the star $\pi$ in mas (milliarcseconds).

The radial velocity is known from observations $V_r$ and two projections of tangential velocity $V_l=4.74r\mu_l\cos b$ and $V_b=4.74r\mu_b,$ directed along galactic longitude $l$ and latitude $b$ respectively, expressed in km/s. Here, the coefficient 4.74 is the ratio of the number of kilometers in an astronomical unit to the number of seconds in a tropical year. Components of proper motion $\mu_l\cos b$ and $\mu_b$ expressed in mas/year.

Through components $V_r, V_l, V_b,$ velocities are calculated $U,V,W,$ where $U$ is the velocity directed from the Sun to the center of the Galaxy, $V$ in the direction of rotation of the Galaxy and $W$~--- to the north galactic pole:
 \begin{equation}
 \begin{array}{lll}
 U=V_r\cos l\cos b-V_l\sin l-V_b\cos l\sin b,\\
 V=V_r\sin l\cos b+V_l\cos l-V_b\sin l\sin b,\\
 W=V_r\sin b                +V_b\cos b.
 \label{UVW}
 \end{array}
 \end{equation}
Thus, velocityes $U,V,W$ directed along the corresponding coordinate axes $x,y,z$.

\subsection{Constructing the Orbits of Stars}
To construct the orbits of stars in a coordinate system rotating around the center of the Galaxy, we use the epicyclic approximation [16]:
 \begin{equation}
 \renewcommand{\arraystretch}{1.8}
 \begin{array}{lll}\displaystyle
 x(t)= x_0+{U_0\over \displaystyle \kappa}\sin(\kappa t)+{\displaystyle V_0\over \displaystyle 2B}(1-\cos(\kappa t)),  \\
 y(t)= y_0+2A \biggl(x_0+{\displaystyle V_0\over\displaystyle 2B}\biggr) t
       -{\displaystyle \Omega_0\over \displaystyle B\kappa} V_0\sin(\kappa t)
       +{\displaystyle 2\Omega_0\over \displaystyle \kappa^2} U_0(1-\cos(\kappa t)),\\
 z(t)= {\displaystyle W_0\over \displaystyle \nu} \sin(\nu t)+z_0\cos(\nu t),
 \label{EQ-Epiciclic}
 \end{array}
 \end{equation}
where $t$~--- time in million years (we proceed from the ratio 1 pc/1 million years = 0.978 km/s), $A$ and $B$~--- Oort constants; $\kappa=\sqrt{-4\Omega_0 B}$~--- epicyclic frequency; $\Omega_0$ is the angular velocity of the galactic rotation of the local rest standard, $\Omega_0=A-B$; and $\nu=\sqrt{4\pi G \rho_0}$~--- is the frequency of vertical vibrations, where $G$ is the gravitational constant, and $\rho_0$~--- stellar density in the solar vicinity.

Options $x_0,y_0,z_0$ and $U_0,V_0,W_0$ in the system of equations (2) denote the modern positions and velocities of stars, respectively. Elevation of the Sun above the galactic plane $h_\odot$ taken equal to 16 pcs according to work [17]. Speeds $U,V,W$ calculate relative to the local standard of rest using the values $(U_\odot,V_\odot,W_\odot)=(11.1,12.2,7.3)$~km/s obtained by Sch\"onrich et al. [18]. We accepted $\rho_0=0.1~M_\odot/$pc$^3$~[19], what gives $\nu=74$~km/s/kpc. We use the following values of the Oort constants: $A=16.9$~km/s/kpc and $B=-13.5$~km/s/kpc, close to modern estimates. An overview of such estimates can be found, for example, in the work [20].

System of equations (2) allows you to calculate the position of the star at each given moment in time. The values of all constants are determined, the solution is found by substituting the moment in time.

\subsection{Analysis of Instantaneous Velocities of Stars}
According to the linear Ogorodnikov--Milne model [21] under the assumption that the peculiar speed of the Sun $(U,V,W)_\odot$  excluded or equal to zero, speed $U,V,W$ can be represented as the following system of linear equations:
\begin{equation}
 \renewcommand{\arraystretch}{2.2}
 \begin{array}{lll}
 \displaystyle
 \qquad
 U={\left(\frac{\partial U}{\partial x}\right)}_0  x+
   {\left(\frac{\partial U}{\partial y}\right)}_0  y+
   {\left(\frac{\partial U}{\partial z}\right)}_0  z, \\
 \displaystyle
 \qquad
 V={\left(\frac{\partial V}{\partial x}\right)}_0  x+
   {\left(\frac{\partial V}{\partial y}\right)}_0  y+
   {\left(\frac{\partial V}{\partial z}\right)}_0  z, \\
 \displaystyle
 \qquad
 W={\left(\frac{\partial W}{\partial x}\right)}_0  x+
   {\left(\frac{\partial W}{\partial y}\right)}_0  y+
   {\left(\frac{\partial W}{\partial z}\right)}_0  z.
 \label{model-1}
 \end{array}
 \end{equation}
Here, the subscript ``0'' means that the derivatives are taken at the origin of the coordinate system. Note that all nine gradients participating in the system of equations (3) can be found graphically.

Using three diagonal gradients from the right-hand sides of the system of equations (3) you can find the parameters of the expansion of the star system $K_{xy}$, $K_{xz}$ and $K_{yz}$ in three corresponding planes:
\begin{equation}
 \begin{array}{lll}
 \displaystyle
 \qquad
  K_{xy}=\left[
   {\left(\frac{\partial U}{\partial x}\right)}_0+
   {\left(\frac{\partial V}{\partial y}\right)}_0 \right]/2, \\
\displaystyle
 \qquad
  K_{xz}=\left[
   {\left(\frac{\partial U}{\partial x}\right)}_0+
   {\left(\frac{\partial W}{\partial z}\right)}_0 \right]/2, \\
\displaystyle
 \qquad
  K_{yz}=\left[
   {\left(\frac{\partial Y}{\partial x}\right)}_0+
   {\left(\frac{\partial W}{\partial z}\right)}_0 \right]/2.
 \label{Kxy}
 \end{array}
 \end{equation}
Odds $K_{xy}$, $K_{xz}$ and $K_{yz}$ which have the dimension of angular velocity, are important for estimating the time
that has passed since the beginning of the expansion of the stellar system.

%%%%%%%%%%%%%%%%%%%%%%%%%%%%%%%%%%%%%%%%%%%%%%%%%%%%%%%%%%%%% t-1
  \begin{table}[p]
  \caption[]{\small Initial data on association stars $\epsilon$ Cha }
  \begin{center}  \label{Table-1}    \small
  \begin{tabular}{|l|r|r|r|r|r|r|}\hline
 Name & $\alpha$ & $\delta$ & $\pi\pm\sigma$ & $\mu_\alpha\cos\delta\pm\sigma$ & $\mu_\delta\pm\sigma$  \\
      & deg & deg & mas  & mas/yr & mas/yr \\\hline

CP$-$68 1388      &164.46&$-69.23$&$ 8.64\pm0.01$ &$-34.81\pm0.01$ &$  3.79\pm0.01$ \\
MP Mus          &200.53&$-69.64$&$10.22\pm0.01$ &$-38.35\pm0.01$ &$-20.00\pm0.01$ \\
RX J1147.7$-$7842 &176.95&$-78.70$&$ 9.91\pm0.02$ &$-41.27\pm0.03$ &$ -3.81\pm0.03$ \\
RX J1149.8$-$7850 &177.38&$-78.85$&$ 9.94\pm0.02$ &$-42.03\pm0.03$ &$ -4.00\pm0.02$ \\
T Cha           &179.31&$-79.36$&$ 9.74\pm0.03$ &$-41.59\pm0.04$ &$ -8.65\pm0.03$ \\
HD 104036       &179.65&$-77.83$&$ 9.54\pm0.03$ &$-41.23\pm0.03$ &$ -7.73\pm0.03$ \\
RX J1159.7$-$7601 &179.93&$-76.02$&$10.09\pm0.01$ &$-41.03\pm0.01$ &$ -6.02\pm0.01$ \\
HD 104467       &180.41&$-78.99$&$10.18\pm0.12$ &$-42.11\pm0.14$ &$ -5.05\pm0.12$ \\
RX J1202.1$-$7853 &180.51&$-78.88$&$10.18\pm0.06$ &$-43.79\pm0.07$ &$ -4.61\pm0.06$ \\
RX J1204.6$-$7731 &181.15&$-77.53$&$ 9.91\pm0.02$ &$-41.45\pm0.02$ &$ -6.22\pm0.02$ \\
RX J1207.7$-$7953 &181.95&$-79.88$&$10.01\pm0.01$ &$-42.10\pm0.01$ &$ -7.24\pm0.01$ \\
HD 105923       &182.91&$-71.18$&$ 9.47\pm0.01$ &$-38.78\pm0.01$ &$ -7.34\pm0.01$ \\
RX J1216.8$-$7753 &184.19&$-77.89$&$ 9.82\pm0.02$ &$-39.98\pm0.02$ &$ -9.05\pm0.02$ \\
RX J1219.7$-$7403 &184.93&$-74.07$&$ 9.92\pm0.01$ &$-40.35\pm0.01$ &$ -9.14\pm0.01$ \\
2MASS J12210499$-$7116493&185.27&$-71.28$&$10.06\pm0.01$&$-40.56\pm0.01$&$ -9.65\pm0.02$ \\
RX J1239.4$-$7502 &189.84&$-75.04$&$ 9.70\pm0.01$ &$-38.30\pm0.01$ &$-12.21\pm0.01$ \\
CD$-$69 1055      &194.61&$-70.48$&$10.51\pm0.01$ &$-41.13\pm0.01$ &$-16.48\pm0.01$ \\
RX J1158.5$-$7754A&179.62&$-77.91$&$ 9.35\pm0.08$ &$-40.95\pm0.09$ &$-10.58\pm0.11$ \\
RX J1150.9$-$7411 &177.69&$-74.19$&$ 9.44\pm0.03$ &$-38.28\pm0.04$ &$ -3.82\pm0.05$ \\
CXOU J115908.2$-$781232 & 179.78 & $-78.21$ & $  9.41\pm0.03 $ & $-38.95\pm0.03$ & $  -5.48\pm0.03$ \\
2MASS J12005517$-$7820296 & 180.23 & $-78.34$ & $  9.73\pm0.04 $ & $-40.75\pm0.04 $ & $  -5.04\pm0.04$ \\
2MASS J12014343$-$7835472 & 180.43 & $-78.60$ & $  9.55\pm0.07 $ & $-41.39\pm0.08 $ & $  -6.21\pm0.08$ \\
CXOU J120152.8$-$781840 & 180.47 & $-78.31$ & $  9.77\pm0.03 $ & $-40.80\pm0.03 $ & $  -7.03\pm0.03$ \\
RX J1202.8$-$7718   & 180.73 & $-77.31$ & $  9.61\pm0.01 $ & $-39.34\pm0.02 $ & $  -6.12\pm0.02$ \\
USNO-B 120144.7$-$781926 & 180.43 & $-78.32$ & $  9.79\pm0.03 $ & $-41.85\pm0.04 $ & $  -6.13\pm0.04$ \\
HD 104237A        & 180.02 & $-78.19$ & $  9.38\pm0.04 $ & $-39.28\pm0.05 $ & $  -5.78\pm0.05$ \\  \hline
 \end{tabular}\end{center}
 \end{table}
%%%%%%%%%%%% t
%%%%%%%%%%%%%%%%%%%%%%%%%%%%%%%%%%%%%%%%%%%%%%%%%%%%%%%%%%%%% t-2
  \begin{table}[p]
  \caption[]{\small
Radial velocities of association stars $\epsilon$ Cha
 }
  \begin{center}  \label{Table-2}    \small
  \begin{tabular}{|l|r|r|r|r|c|c|}\hline
 Name & $\alpha$ ~~ & $\delta$ ~~ &  $V_r\pm\sigma$ ~~~~&  $(V_r\pm\sigma)_{Gaia DR3}$\\
 & deg & deg & km/s ~~~~ & km/s ~~~~ \\\hline

CP$-$68 1388            & 164.46 & $-69.23$ & $ 15.90\pm1.00 $ & $ 15.69\pm0.63 $ \\
MP Mus                  & 200.53 & $-69.64$ & $ 11.60\pm0.20 $ & $  7.20\pm1.65 $ \\
RX J1147.7$-$7842       & 176.95 & $-78.70$ & $ 16.10\pm0.90 $ & $ 18.52\pm3.40 $ \\
RX J1149.8$-$7850       & 177.38 & $-78.85$ & $ 13.40\pm1.30 $ & $ 14.56\pm1.60 $ \\
T Cha                   & 179.31 & $-79.36$ & $ 18.00\pm2.00 $ & $ 17.31\pm2.63 $ \\
HD 104036               & 179.65 & $-77.83$ & $ 12.60\pm0.50 $ & $ 13.55\pm0.61 $ \\
RX J1159.7$-$7601       & 179.93 & $-76.02$ & $ 13.00\pm3.70 $ & $ 12.68\pm0.61 $ \\
HD 104467               & 180.41 & $-78.99$ & $ 12.81\pm0.96 $ & $ 13.10\pm0.81 $ \\
RX J1202.1$-$7853       & 180.51 & $-78.88$ & $ 17.10\pm0.20 $ & $ 13.91\pm1.62 $ \\
RX J1204.6$-$7731       & 181.15 & $-77.53$ & $ 10.40\pm2.00 $ & $ 12.96\pm4.45 $ \\
RX J1207.7$-$7953       & 181.95 & $-79.88$ & $ 15.00\pm0.70 $ & $ 13.71\pm2.77 $ \\
HD 105923               & 182.91 & $-71.18$ & $ 14.34\pm1.06 $ & $ 14.26\pm0.69 $ \\
RX J1216.8$-$7753       & 184.19 & $-77.89$ & $ 14.00\pm2.00 $ & $ 13.28\pm7.23 $ \\
RX J1219.7$-$7403       & 184.93 & $-74.07$ & $ 13.86\pm1.89 $ & $ 13.74\pm1.49 $ \\
2MASS J12210499$-$71164 & 185.27 & $-71.28$ & $ 11.44\pm2.53 $ & $ 11.56\pm0.68 $ \\
RX J1239.4$-$7502       & 189.84 & $-75.04$ & $ 13.62\pm2.80 $ & $ 12.66\pm0.59 $ \\
CD$-$69 1055            & 194.61 & $-70.48$ & $ 11.18\pm1.67 $ & $ 10.68\pm1.00 $ \\
RX J1158.5$-$7754A      & 179.62 & $-77.91$ & $ 14.02\pm1.82 $ & $ 10.25\pm1.26 $ \\
RX J1150.9$-$7411       & 177.69 & $-74.19$ & $ 15.00\pm1.20 $ & $    $ \\
CXOU J115908.2$-$781232 & 179.78 & $-78.21$ & $ 15.10\pm0.20 $ & $    $ \\
2MASS J12005517$-$78202 & 180.23 & $-78.34$ & $ 10.70\pm1.30 $ & $    $ \\
2MASS J12014343$-$78354 & 180.43 & $-78.60$ & $ 20.00\pm0.60 $ & $    $ \\
CXOU J120152.8$-$781840 & 180.47 & $-78.31$ & $ 16.50\pm1.10 $ & $    $ \\
RX J1202.8$-$7718       & 180.73 & $-77.31$ & $ 14.40\pm0.60 $ & $    $ \\
USNO-B 120144.7$-$78192 & 180.43 & $-78.32$ & $ 14.90\pm1.10 $ & $    $ \\
HD 104237A              & 180.02 & $-78.19$ & $ 13.52\pm0.39 $ & $    $ \\\hline
 \end{tabular}\end{center}
  \end{table}
%%%%%%%%%%%% t2
%%%%%%%%%%%%%%%%%%%%%%%%%%%%%%%%%%%%%%%%%%%%%%%%%%%%%%%%%%%%% F1:
\begin{figure}[t]
{ \begin{center}
  \includegraphics[width=0.9\textwidth]{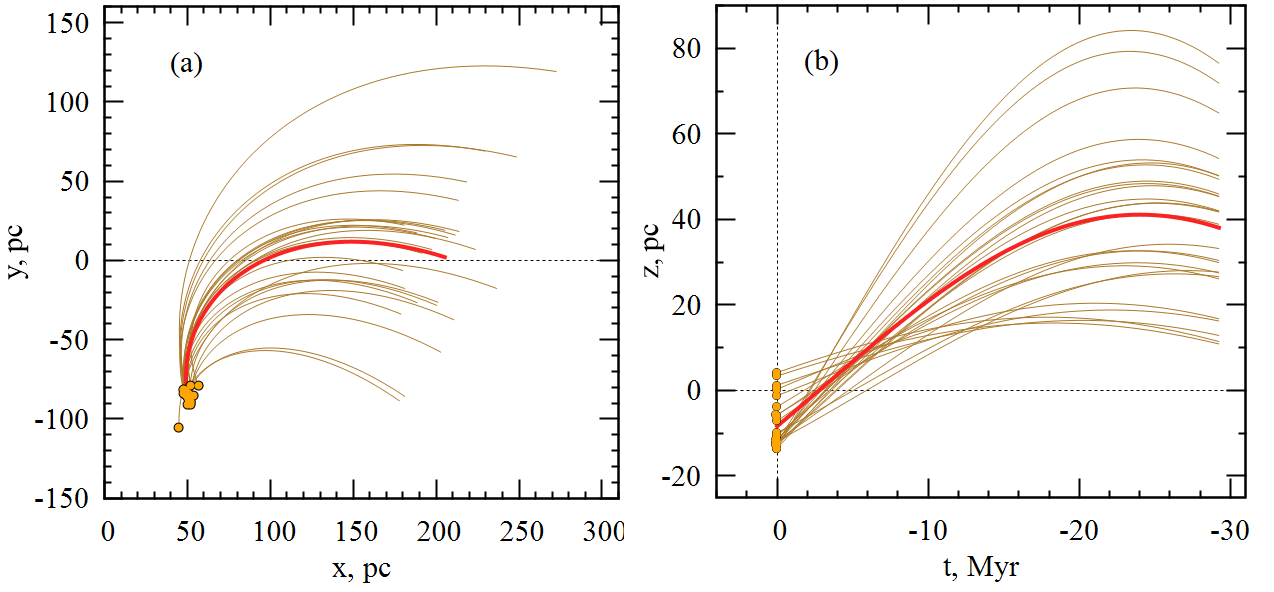}
  \caption{
of 26 association stars $\epsilon$~Cha in projection onto the galactic  plane $xy$ and their trajectories (a), vertical distribution
and their trajectories traced back into the past over an interval of 30 million years (b). The trajectory of the kinematic center
of the association is shown in red.}
 \label{f1-DR3}
\end{center}}
\end{figure}
%%%%%%%%%%%%%%%%%%%%  f1
%%%%%%%%%%%%%%%%%%%%%%%%%%%%%%%%%%%%%%%%%%%%%%%%%%%%%%%  F2
\begin{figure}[t]
{ \begin{center}
   \includegraphics[width=0.7\textwidth]{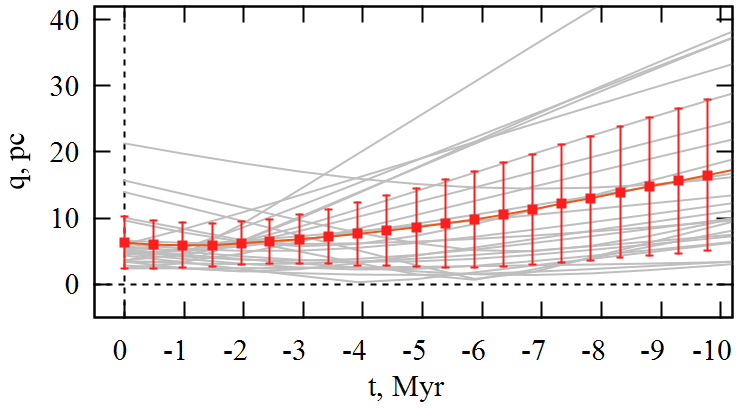}
  \caption{
Deviations from the trajectory of the kinematic center (parameter $q$) on an integration interval of 10 million years for 26
stars of the association $\epsilon$~Cha. Average values and corresponding variances are shown in red.} 
\label{f2-q}
\end{center}}
\end{figure}
%%%%%%%%%%%%%%%%%%%%  f2
%%%%%%%%%%%%%%%%%%%%%%%%%%%%%%%%%%%%%%%%%%%%%%%%%%%%%%%%%%%%% F3:
\begin{figure}[t]
{ \begin{center}
   \includegraphics[width=0.99\textwidth]{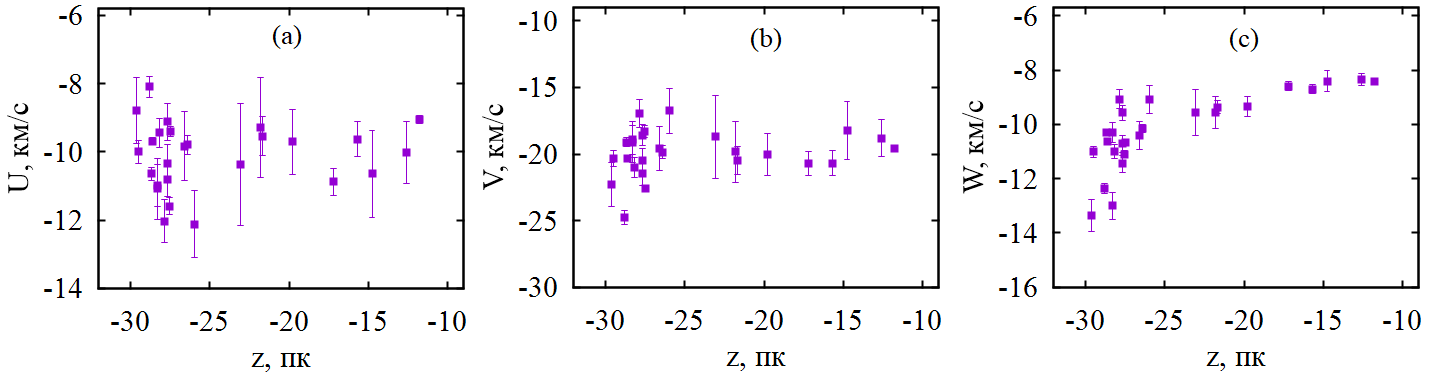}
  \caption{
Speed dependencies $U$~(a),  $V$~(b) and $W$~(c) from coordinate $z$ for 26 stars of the association $\epsilon$~Cha.}
 \label{f3-UVW-Z}
\end{center}}
\end{figure}
%%%%%%%%%%%%%%%%%%%%  f3
%%%%%%%%%%%%%%%%%%%%%%%%%%%%%%%%%%%%%%%%%%%%%%%%%%%%%%%%%%%%% F4:
\begin{figure}[t]
{ \begin{center}
   \includegraphics[width=0.98\textwidth]{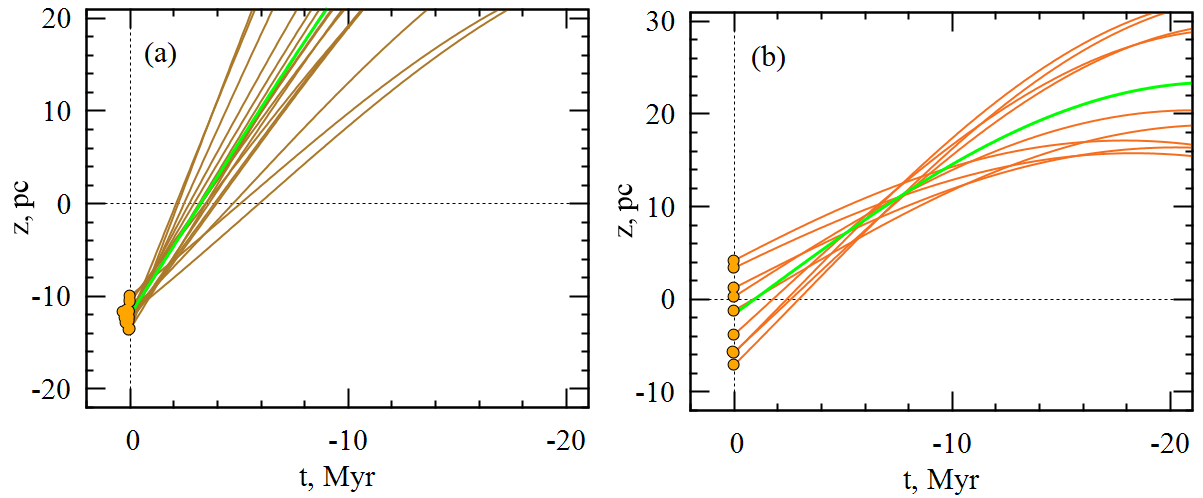}
  \caption{
Vertical distribution of 17 stars with $z<-25$~pc and their trajectories~(a), 9 stars with $z>-25$~pc and their trajectories
(b). The trajectories of the kinematic centers, which are calculated from the stars of these samples, are shown in green.}
 \label{f4-UVW-Z}
\end{center}}
\end{figure}
%%%%%%%%%%%%%%%%%%%%  f4
%%%%%%%%%%%%%%%%%%%%%%%%%%%%%%%%%%%%%%%%%%%%%%%%%%%%%%%%%%%%% F5:
\begin{figure}[t]
{ \begin{center}
   \includegraphics[width=0.99\textwidth]{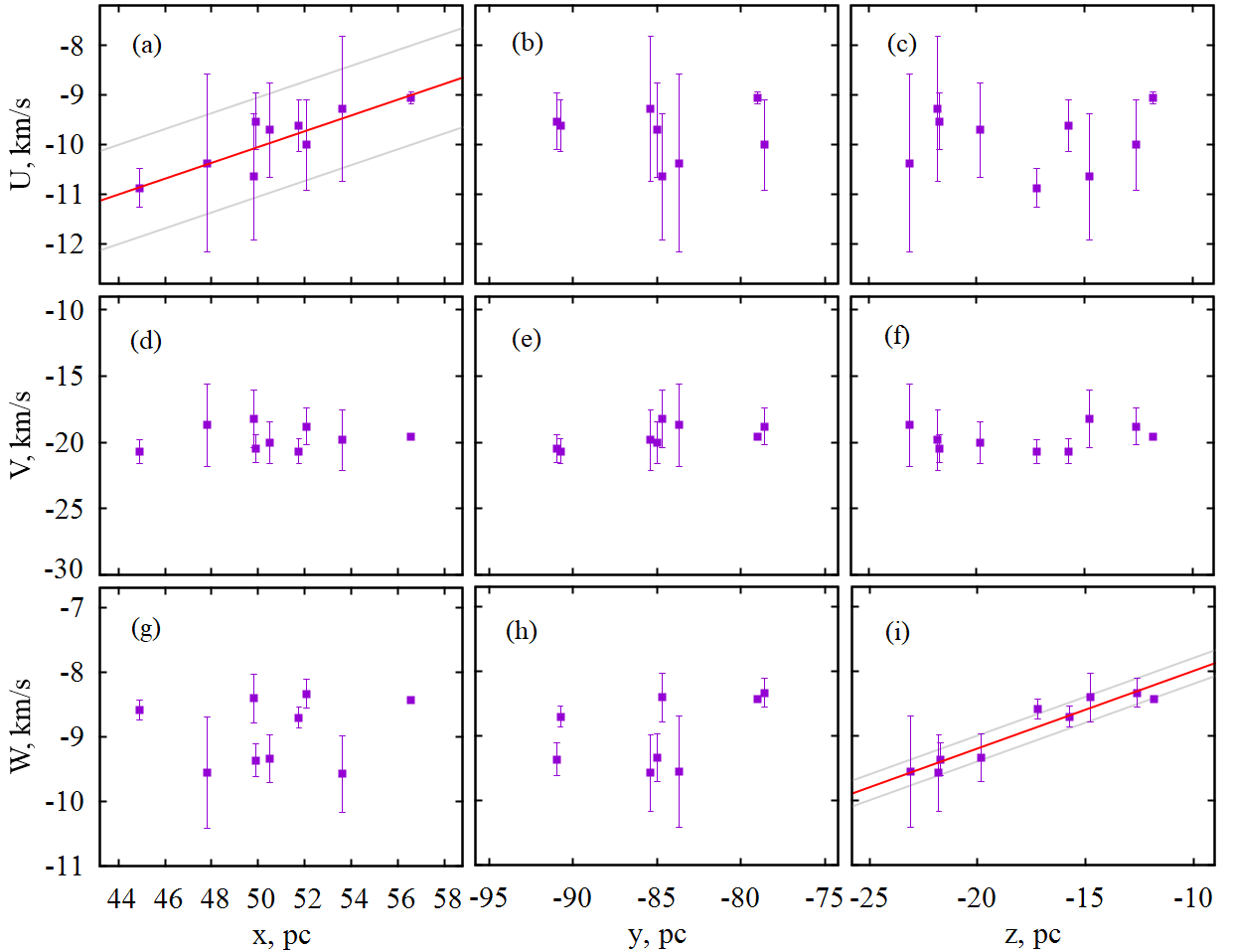}
  \caption{
Speed dependencies $U,V,W$ from coordinates $x,y,z$ for a sample of 9 stars with $z>-25$~pc~(b).}
 \label{f5-UVW-xyz-9}
\end{center}}
\end{figure}
%%%%%%%%%%%%%%%%%%%%  f5
%%%%%%%%%%%%%%%%%%%%%%%%%%%%%%%%%%%%%%%%%%%%%%%%%%%%%%%%%%%%% F6:
\begin{figure}[t]
{ \begin{center}
   \includegraphics[width=0.99\textwidth]{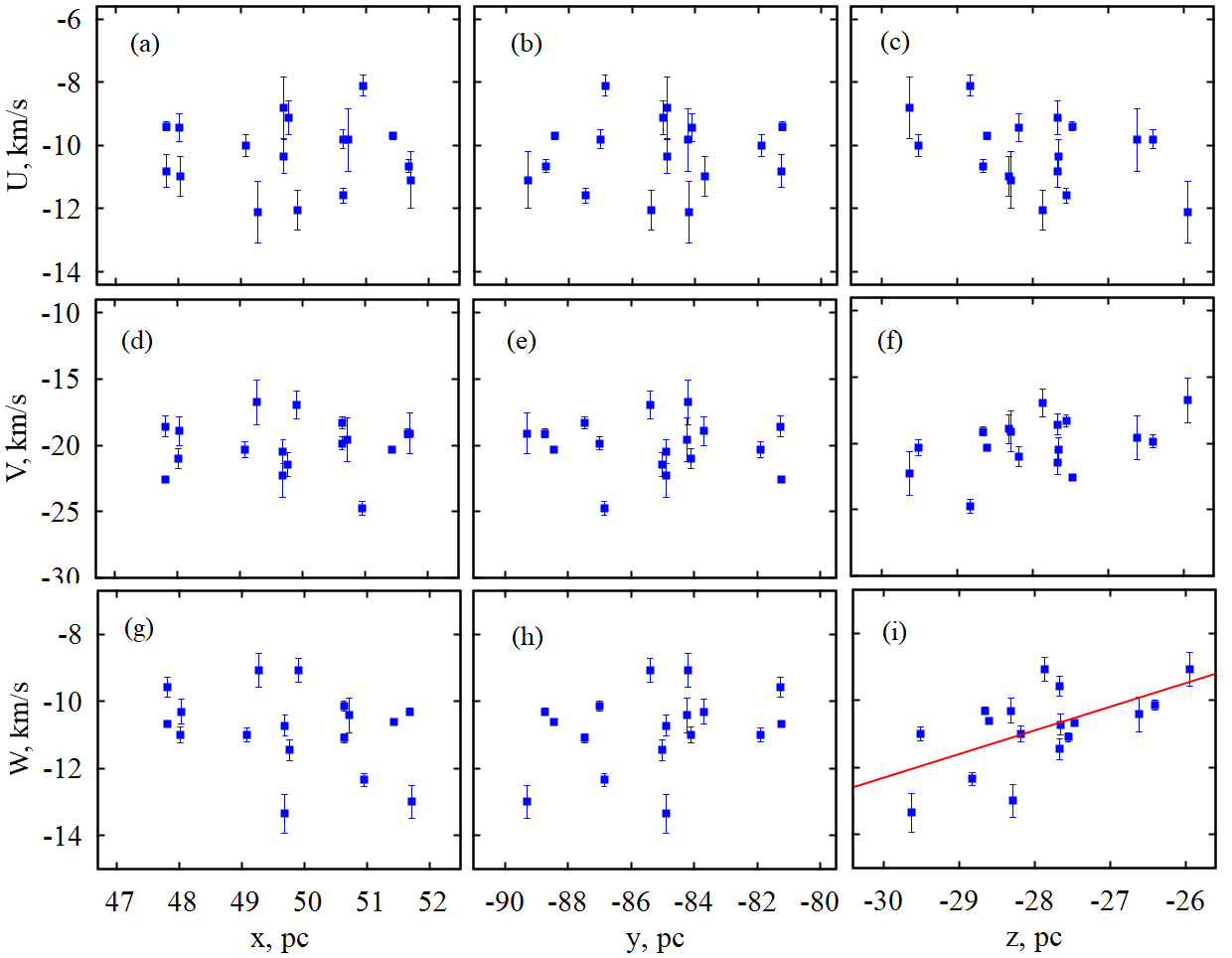}
  \caption{
Speed dependencies $U,V,W$ from coordinates $x,y,z$ for a sample of 17 stars with $z<-25$~pc~(b).}
 \label{f6-UVW-xyz-17}
\end{center}}
\end{figure}
%%%%%%%%%%%%%%%%%%%%  f6
%%%%%%%%%%%%%%%%%%%%%%%%%%%%%%%%%%%%%%%%%%%%%%%%%%%%%%%%%%%%% F7:
\begin{figure}[t]
{ \begin{center}
   \includegraphics[width=0.98\textwidth]{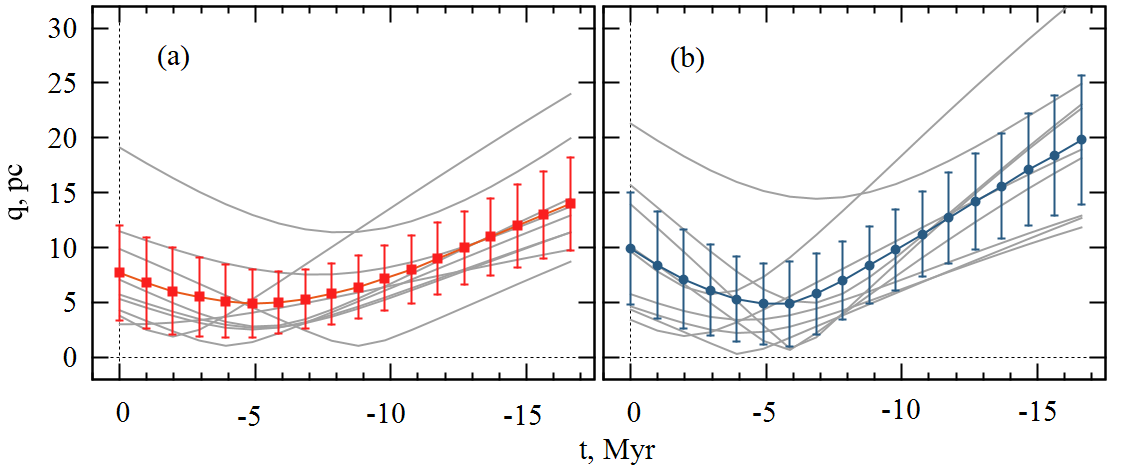}
  \caption{
Parameter dependency $q$ from time for a sample of 9 stars with $z>-25$~pc whose orbits were constructed using the kinematic center calculated from these 9 stars (a), and with the kinematic center parameters obtained from all 26 stars of the association (b).}
 \label{f7-sred-9}
\end{center}}
\end{figure}
%%%%%%%%%%%%%%%%%%%%  f7
%%%%%%%%%%%%%%%%%%%%%%%%%%%%%%%%%%%%%%%%%%%%%%%%%%%%%%%%%%%%% F8:
\begin{figure}[t]
{ \begin{center}
   \includegraphics[width=0.98\textwidth]{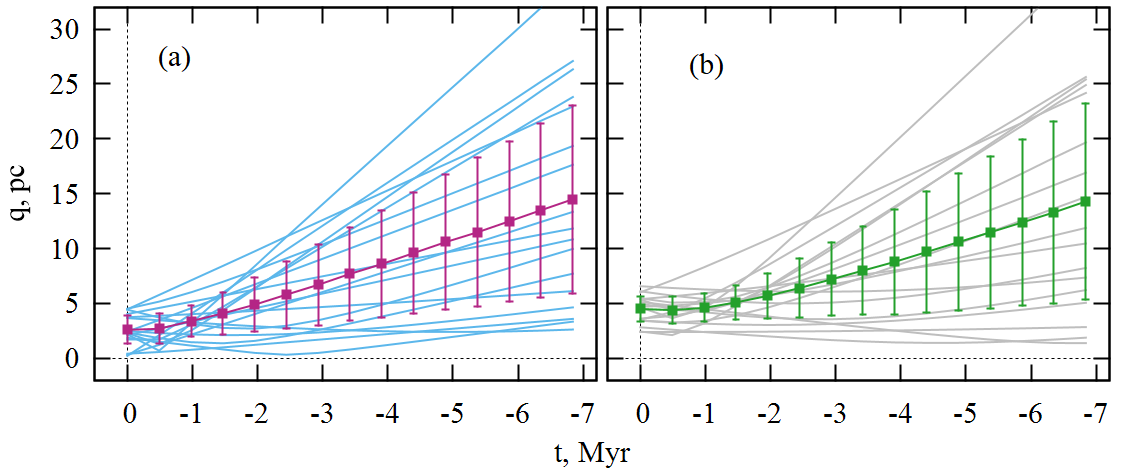}
  \caption{
Parameter dependency $q$ from time for a sample of 17 stars with $z<-25$~pc whose orbits were constructed using the kinematic center calculated from these 17 stars (a), and with the kinematic center parameters obtained from all 26 stars of the association (b).}
 \label{f8-sred-17}
\end{center}}
\end{figure}
%%%%%%%%%%%%%%%%%%%%  f8

\section{RESULTS AND DISCUSSION}
Our sample is based on association stars $\epsilon$~Cha from the list [1]. These authors studied the kinematic properties of the association using data from the Gaia~DR2 catalog. In this work, we supplied stars from the list [1] data from Gaia~DR3 version [14]. Note that the sample did not include several binary systems from the list [1], for which there are no measurements in the catalog [14]. Thus, our working sample contains 26 stars, marked with the index ``M'' (member) in the work [1].

Initial data on the selected 26 stars of the association $\epsilon$~Cha are contained in Tables 1 and 2. As can be seen from Table 2, radial velocities from the Gaia~DR3 catalog, usually measured from larger errors compared to ground-based measurements are not available for all selected stars. Bibliographic references to ground-based measurements of radial velocities are given in the work [1].

In Fig. 1, the current positions of the 26 stars of the association are given $\epsilon$~Cha and their trajectories, built back to the past. The trajectory of the kinematic center of the association is also given. To set the trajectory of the kinematic center, the average positions and velocities
of the stars were calculated ${\overline x}_0,{\overline y}_0,{\overline z}_0$ and ${\overline U}_0,{\overline V}_0,{\overline W}_0$. As a result, the following values were calculated for 26 stars: ${\overline x}_0=50.2$~pc, ${\overline y}_0=-85.8$~pc and ${\overline z}_0=-24.4$~pc, and also
${\overline U}_0=-10.11$~km/s, ${\overline V}_0=-19.90$~km/s and ${\overline W}_0=-10.18$~km/s.

Using differences (between the star and the kinematic center) of coordinates $\Delta x,\Delta y,\Delta z$ at each moment of integration for each star we calculate the value of the parameter of the following form:
 \begin{equation}
 \renewcommand{\arraystretch}{1.8}
 q=\sqrt{\Delta x^2+\Delta y^2+\Delta z^2},
 \label{qq}
 \end{equation}
which characterizes the deviation of the star from the trajectory of the kinematic center. Note that the trajectories of stars were calculated taking into account the elevation of the Sun above the galactic plane. Thus, in all our drawings (with star trajectories), the coordinate $z$ reflects the position of stars relative to the Galactic plane.

In Fig. 2 for each star, the dependence of the parameter is given $q$ from time over an integration interval of 10 million years. It can be seen that the size of the star group 10 million years in the past had a significantly larger spatial size compared to the present. However, based on the position of the minimum of the red curve, we would have to conclude that the kinematic age of the association $\epsilon$~Cha is about 1 million years old, which is very different from known estimates obtained by other methods.

In Fig. 3, speed dependences are given $U,$ $V$ and $W$  from coordinate $z$ for 26 stars of the association $\epsilon$~Cha. In all three panels, two clusters of stars are clearly visible, which are separated by a boundary $z=-25$~pc. They are especially clearly visible in Fig. 3(c). 

Therefore, it was further decided to analyze each condensation separately. Using just one boundary condition $z=-25$~pc two samples were obtained. Sample selected under the condition $z>-25$~pc contains 9 stars, and the selected one provided $z<-25$~pc contains 17 stars. As it turns out, these samples have very different kinematic properties.

Given that $z>-25$~pc the following 9 stars were selected:
 CP$-$68\,1388,
 MP\,Mus,
 RX\,J1159.7$-$7601,
 HD\,105923,
 RX\,J1219.7$-$7403,
 2MASS\,J12210499$-$7116493,
 RX\,J1239.4$-$7502,
 CD$-$69 1055 and
 RX\,J1150.9$-$7411.

And on condition $z<-25$~pc, the following 17 stars were selected:
RX\,J1147.7$-$7842,
RX\,J1149.8$-$7850,
T\,Cha,
HD\,104036,
HD\,104467,
RX\,J1202.1$-$7853,
RX\,J1204.6$-$7731,
RX\,J1207.7$-$7953,
RX\,J1216.8$-$7753,
RX\,J1158.5$-$7754A,
CXOU\,J115908.2$-$781232,
2MASS\,J12005517$-$78202,
2MASS\,J12014343$-$78354,
CXOU\,J120152.8$-$781840,
RX\,J1202.8$-$7718,
USNO-B\,120144.7$-$78192 and
HD\,104237A.

In Fig. 4, the vertical distribution of stars in two samples and the trajectories of their kinematic centers, calculated from the stars of each sample, are given. Panel (a) clearly shows that the stars were very recently ($\sim$1~million years ago) came together, but the stars in panel (b) occupied a minimum size, at least along the axis $z$ about 8 million years ago, and are currently showing expansion.

In Fig. 5, speed dependences are given $U,$ $V,$ $W$ from coordinates $x,y,z$ for a sample of 9 stars selected under the condition $z>-25$~pc. In panels (a) and (I), the red lines show the dependences found from these data using the least squares method, corresponding to the gradients $\partial U/\partial x=159\pm36$~km/s/kpc and $\partial W/\partial z=120\pm14$~km/s/kpc. The boundaries of the trust areas corresponding to the level are indicated $1\sigma$. No other gradients significantly different from zero were found for these stars. Based on the ratio (4) using the two found gradients, we can estimate the coefficient of angular velocity of expansion of this stellar grouping in the plane $xz$:
 \begin{equation}
 K_{xz}=135\pm19~\hbox {km/s/kpc}
 \label{t-xz}
 \end{equation}
and find the period of time that has passed from the beginning of the expansion of the stellar system to the present moment, $t=977.5/K_{xz}$:
 \begin{equation}
 t=7.2\pm1.0~\hbox {Myr.}
 \label{t-Kxz-9}
 \end{equation}

In Fig. 6, speed dependences are given $U,$ $V,$ $W$ from coordinates $x,y,z$  for a sample of 17 stars selected under the condition $z<-25$~pc. Using the least squares method, only one gradient significantly different from zero was found for the stars in this sample,
$\partial W/\partial z=707\pm248$~km/s/kpc. The corresponding dependence is given by the red line in Fig. 6i, where the confidence region is not shown because it is too wide. The corresponding value of this gradient is the time elapsed from the beginning of expansion
along the axis $z$ star system up to the present day is:
\begin{equation}
 t=1.4\pm0.5~\hbox {Myr.}
 \label{t-dWdz-17}
 \end{equation}

In Fig. 7, the orbits of a sample of 9 stars are given with $z>-25$~pcs built for two variants of the kinematic center. Fig. 7a uses the parameters of the kinematic center calculated from 9 sample stars. In Fig. 7b, the parameters of the kinematic center were used, calculated for all 26 stars of the association. In both cases, the deflection of the line of average values in the region of 5 million years is clearly visible, and in the case of Fig. 7b, the deflection is more pronounced. However, panel (a) shows smaller variances in each interval. Analysis of the average values and their variances
shown in the figure makes it possible to estimate the point in time at which the size of the stellar system in the past was minimal. As a result, the value was found
 \begin{equation}
 t=4.9\pm0.8~\hbox {Myr.}
 \label{t-9}
 \end{equation}
Here, the mistake of the moment $t$ was found as a result of statistical modeling using the Monte Carlo method. It was assumed that the orbits of the stars were constructed with errors of 10\%, distributed according to the normal law.

In Fig. 8, the orbits of a sample of 17 stars are given with $z<-25$~pc. From the analysis of the average values and their variances shown in the figure, we obtained the value of the time interval that has passed since the moment when the size of the system was minimal,
\begin{equation}
 t=0.2\pm0.3~\hbox {Myr.}
 \label{t-17}
 \end{equation}
The same as when calculating the result (9), torque error was found as a result of statistical modeling using the Monte Carlo method under the assumption that the orbits of stars are constructed with errors of 10\%, distributed according to a normal law. This grouping of young stars is of greatest interest, since the estimates found for them (8) and (10) differ greatly from the results of estimates of the characteristic age of the association $\epsilon$~Cha, 5--7 million years, obtained by other authors (noted by us in the Introduction).

\section{CONCLUSIONS}
A sample of probable members of the association was studied $\epsilon$~Cha. Based on a list of candidate stars
from the work, [1] a working sample of 26 stars was compiled. These stars are provided with trigonometric
parallaxes, proper motions from the Gaia~DR3 catalog, and radial velocities taken from literature sources.

To estimate the kinematic age of the association $\epsilon$~Cha were: (a) the orbits of stars back into the past were
constructed, and the moment when the stellar grouping had a minimum spatial size was determined; and
(b) the dependences of the observed velocities are plotted $U,V,W$ from coordinates $x,y,z$ in order to
identify the expansion effect. This made it possible to obtain association age estimates in two ways, both of which are kinematic.

When analyzing stellar trajectories and the dependences of velocities on coordinates, it was found that
the original sample is divided into two parts. As it turned out, both parts have different kinematic properties.

The first sample included 9 stars. Based on constructing the orbits of stars back into the past, an age estimate was obtained for these stars $t=4.9\pm0.8$~million years (result (9)). Speed dependency analysis $U,V,W$ from coordinates $x,y,z$  showed the presence
of expansion in the plane  $xz$ with magnitude $K_{xz}=135\pm19$~km/s/kpc, on the basis of which the
time interval elapsed from the beginning of the expansion of the stellar system to the present moment was
found $t=7.2\pm1.0$~million years (result (7)). We see that both estimates agree well with each other within the limits of the errors found.

The second sample included 17 stars. Based on constructing the orbits of stars back into the past, an age estimate was obtained for these stars
$t=0.2\pm0.3$~million years (result (10)). Speed dependency analysis $U,V,W$ from coordinates $x,y,z$ showed the presence of one gradient that describes the expansion along the axis$z$ with magnitude $\partial W/\partial z=707\pm248$~km/s/kpc, on the basis of which the time interval
elapsed from the beginning of the expansion of this part of the association was found $\epsilon$~Cha until now $t=1.4\pm0.5$~million years (result (8)). Here too, we see good agreement between the estimates obtained by the two methods.

Our sample uses the coordinates and velocities of stars measured with high accuracy. Moreover, with a difference of several $\sigma$, different age estimates were obtained for the two groups. Two scenarios can be proposed to explain the results obtained. In the first case, it seems possible that the association $\epsilon$~Cha consists of two independent parts of different ages. In the second option, both groups have a common long-term evolution, but the younger one could have arisen as a result of a recent burst of star formation. Both of these scenarios do not contradict the fact that stellar associations have a complex hierarchical structure [22, 23].

Note, for example, that in the early stages of formation, star clusters can undergo rapid dynamic evolution, leading to strong gravitational influences and the ejection of ``escaping'' stars. The processes of star loss in the vicinity of the star cluster NGC 1976 (ONC) are considered in the work [24]. In progress, [25] using calculations of the spatial motion of stars in past epochs, the discovery of an ``old'' stellar ``relic filament'' associated with the star formation region in Orion was reported.

Calculations of the spatial movements of the gas clouds Orion A and Orion B showed that they were located next to each other about 6 million years ago, and are now moving away radially from approximately the same region of space [26]. In progress [27] published a study of the three-dimensional structure, kinematics and age distribution of the Orion association based on Gaia~DR2 data.

\subsubsection*{ACKNOWLEDGMENTS}
The authors are grateful to the reviewer for useful comments that contributed to the improvement of the work.

\subsubsection*{FUNDING}
This work was supported by ongoing institutional funding. No additional grants to carry out or direct this particular
research were obtained.

\subsubsection*{CONFLICT OF INTEREST}
The authors of this work declare that they have no conflicts of interest.

 \subsubsection*{REFERENCES}
 \small

\quad~~1. D. A. Dickson-Vandervelde, E. C. Wilson, and J. H. Kastner, Astron. J. 161, 87 (2021).

2. E. E. Mamajek, W. A. Lawson, and E. D. Feigelson, Astrophys. J. 544, 356 (2000).

3. E. D. Feigelson, W. A. Lawson, and G. P. Garmire, Astrophys. J. 599, 1207 (2003).

4. K. L. Luhman, Astrophys. J. 616, 1033 (2004).

5. E. Jilinski, V. G. Ortega, and R. de la Reza, Astrophys. J. 619, 945 (2005).

6. C. A. O. Torres, G. R. Quast, C. H. F. Melo, and M. F. Sterzik, in Handbook of Star Forming Regions,
Vol. 2: The Southern Sky, Vol. 5 of ASP Monograph Publications, Ed. by Bo Reipurth (2008), p. 757.

7. B. Lopez Marti, F. Jimenez Esteban, A. Bayo, D. Barrado, E. Solano, and C. Rodrigo, Astron. Astrophys.
551, A46 (2013).

8. C. Brice\~no and A. Tokovinin, Astron. J. 154, 195 (2017).

9. K. Kubiak, K. Mu\v zi\'c, I. Sousa, V. Almendros-Abad, R. Kohler, and A. Scholz, Astron. Astrophys. 650, A48
(2021).

10. A.-R. Lyo, W. A. Lawson, and M. S. Bessell, Mon. Not. R. Astron. Soc. 389, 1461 (2008).

11. L. L. Kiss, A. Mo\'or, T. Szalai, J. Kov\'acs, et al., Mon. Not. R. Astron. Soc. 411, 117 (2011).

12. M. Fang, R. van Boekel, J. Bouwman, T. Henning, W. A. Lawson, and A. Sicilia-Aguilar, Astron. Astrophys.
549, A15 (2013).

13. A. G. A. Brown, A. Vallenari, T. Prusti, J. H. J. de Bruijne, et al., Astron. Astrophys. 616, A1 (2018).

14. A. Vallenari, A. G. A. Brown, T. Prusti, J. H. J. de Bruijne, et al., Astron. Astrophys. 674, A1 (2023).

15. A. Kunder, G. Kordopatis, M. Steinmetz, T. Zwitter, et al., Astron. J. 153, 75 (2017).

16. B. Lindblad, Ark. Mat. Astron. Fys. 20A (17) (1927).

17. V. V. Bobylev and A. T. Bajkova, Astron. Lett. 42, 1 (2016).

18. R. Sch\"onrich, J. Binney, and W. Dehnen, Mon. Not. R. Astron. Soc. 403, 1829 (2010).

19. J. Holmberg and C. Flynn, Mon. Not. R. Astron. Soc. 352, 440 (2004).

20. O. I. Krisanova, V. V. Bobylev, and A. T. Bajkova, Astron. Lett. 46, 370 (2020).

21. K. F. Ogorodnikov, {\it Dynamics of stellar systems}, Ed. by A. Beer (Pergamon, Oxford, 1965).

22. S. Ratzenb\"ock, J. E. Gro{\ss}schedl, T. M\"oller, J. Alves, I. Bomze, and S. Meingast, Astron. Astrophys.
677, A59 (2023).

23. S. Ratzenb\"ock, J. E. Gro{\ss}schedl, J. Alves, N. Miret-Roig, et al., Astron. Astrophys. 678, A71 (2023).

24. M. Fajrin, J. J. Armstrong, J. C. Tan, J. Farias, and L. Eyer, arXiv: 2402.12258 [astro-ph.SR] (2024).

25. T. Jerabkova, H. M. J. Boffin, G. Beccari, and R. I. Anderson, Mon. Not. R. Astron. Soc. 489, 4418 (2019).

26. J. E. Gro{\ss}schedl, J. Alves, S. Meingast, and G. Herbst-Kiss, Astron. Astrophys. 647, A91 (2021).

27. J. A. Caballero, A. de Burgos, F. J. Alonso-Floriano, A. Cabrera-Lavers, D. Garcia-Alvarez, and D. Montes,
Astron. Astrophys. 629, A114 (2019).

 \end{document}